\documentclass[a4paper,11pt]{article}

\usepackage[T1]{fontenc}
\usepackage{lmodern}
\usepackage[margin=1in]{geometry}
\usepackage{amsmath}
\usepackage{amssymb}
\usepackage{mathtools}
\usepackage{bm}
\usepackage{slashed}
\usepackage{booktabs}

\usepackage{tocloft}
\usepackage[colorlinks=true,linkcolor=black,citecolor=blue,urlcolor=blue]{hyperref}
\usepackage[
  backend=biber,
  style=numeric-comp,
  sorting=none,
  giveninits=true,
  maxnames=99,
  doi=false,
  eprint=false,
  url=false
]{biblatex}
\DeclareNameAlias{author}{given-family}
\DeclareFieldFormat{plainbibtitle}{#1}
\DeclareFieldFormat{plainbibpages}{#1}
\DeclareFieldFormat{plainbibeprint}{#1}
\newbibmacro*{linkedjournal}{%
  \mkbibemph{\printfield{journaltitle}}\space
  \mkbibbold{\printfield{volume}}\space
  (\printfield{year})\space
  \printfield[plainbibpages]{pages}%
}
\DeclareBibliographyDriver{article}{%
  \usebibmacro{bibindex}%
  \usebibmacro{begentry}%
  \printnames{author}%
  \setunit{\addcomma\space}%
  \printtext{\mkbibemph{\printfield[plainbibtitle]{title}}}%
  \iffieldundef{journaltitle}
    {\iffieldundef{eprint}
      {\iffieldundef{year}{}{\setunit{\addcomma\space}\printfield{year}}}
      {}}
    {\setunit{\addcomma\space}%
     \iffieldundef{doi}
       {\usebibmacro{linkedjournal}}
       {\href{https://doi.org/\thefield{doi}}{\usebibmacro{linkedjournal}}}%
    }%
  \iffieldundef{eprint}{}{%
    \setunit{\addcomma\space}%
    \printtext{[}%
    \href{https://arxiv.org/abs/\thefield{eprint}}{\printfield[plainbibeprint]{eprint}}%
    \printtext{]}%
  }%
  \usebibmacro{finentry}%
}

\hypersetup{
  pdftitle={A local second law for black hole dynamics in the charged large D membrane paradigm},
  pdfauthor={Supratim Halder and Mangesh Mandlik}
}

\renewcommand{\footnoterule}{\kern -3pt\hrule width 2.05in\kern 2.6pt}

\newcommand{\dd}{\mathrm{d}}

\begin{document}
\thispagestyle{plain}
\begin{center}
\vspace*{0.23in}
{\fontsize{18}{21}\selectfont\bfseries
A local second law for black hole dynamics in the charged
large $D$ membrane paradigm\par}
\vspace{0.53in}
{\fontsize{14}{18}\selectfont\bfseries
Supratim Halder\textsuperscript{1} and
Mangesh Mandlik\textsuperscript{2}\par}
\vspace{0.23in}
{\fontsize{12}{15}\selectfont\itshape
Department of Physics, Indian Institute of Technology\\
(Indian School of Mines) Dhanbad, Jharkhand 826004, India\par}
\vspace{1.05in}
{\bfseries Abstract\par}
\vspace{0.15in}
\end{center}
\begin{center}
\begin{minipage}{0.87\textwidth}
\noindent
The large $D$ membrane paradigm provides a perturbative, non-gravitational description of black hole dynamics. While the local second law of thermodynamics has been established for neutral membranes, the explicit determination of the local entropy production in the charged large $D$ membrane paradigm remains an open problem. In this paper, we compute the subleading-order correction to the divergence of the membrane velocity for a charged black hole in a flat background. Using the non-affine null Raychaudhuri equation and the
Einstein-Maxwell equation at the physical event horizon, we relate the horizon expansion to the auxiliary membrane kinematics. We demonstrate that the membrane velocity produces a non-negative divergence at the first non-vanishing order in the membrane worldvolume, which is driven by two quadratic forms: the membrane shear and an additional term arising from the Maxwell field. Using this result, we subsequently construct the corresponding membrane entropy current and establish the non-negativity of its divergence through this order, thereby establishing the local second law for the charged large $D$ membrane paradigm.
\end{minipage}
\end{center}
\footnotetext[1]{\href{mailto:23dr0184@iitism.ac.in}{23dr0184@iitism.ac.in}}
\footnotetext[2]{\href{mailto:mandlik@iitism.ac.in}{mandlik@iitism.ac.in}}
\clearpage
\vspace*{0.02in}
\tableofcontents
\vspace{0.9em}
\flushbottom

\section{Introduction}
\label{sec:introduction}
The large $D$ limit of general relativity provides a perturbative framework for studying the nonlinear dynamics of black holes
\cite{Emparan:2013moa,Emparan:2014aba,Emparan:2015rva}.
As the spacetime dimension $D$ increases, the gravitational field becomes
localized within a region of thickness $\mathcal O(r_0/D)$ near the
event horizon, where $r_0$ is the characteristic radius of the black hole. The quasinormal mode spectrum separates into two different scales, one
with frequencies of order $1/r_0$ and another with
frequencies of order $D/r_0$. This separation allows the slow dynamics
to be described by an effective theory. Consequently, the membrane provides a non-gravitational effective description of the nonlinear, long-wavelength dynamics of the black-hole geometry~\cite{Bhattacharyya:2017hpj,Kundu:2018dvx,Dandekar:2016fvw}. These dynamics are encoded in the shape and velocity of a
codimension-one timelike hypersurface embedded in an auxiliary
background spacetime
\cite{Bhattacharyya:2015dva,Bhattacharyya:2015fdk,Dandekar:2016fvw}. For charged black holes, the membrane also carries a charge field, and its evolution follows from the Einstein--Maxwell equations
\cite{Bhattacharyya:2015fdk}. These equations are solved in a perturbative series. In this formulation, to ensure the regularity of the next perturbative order, some gravitational constraint equations have to be obeyed. These are called membrane equations; they dictate the profile of membrane quantities (shape, membrane velocity and charge density).  Those equations were initially introduced in \cite{Bhattacharyya:2015dva,Bhattacharyya:2015fdk}, building upon the earlier foundational work \cite{Emparan:2013moa,Emparan:2013xia,Emparan:2013oza,Emparan:2014cia,Emparan:2014jca,Emparan:2014aba,Emparan:2015rva}. For the stationary solutions, alternative derivations of the membrane dynamics are presented in refs.~\cite{Emparan:2015hwa,Suzuki:2015iha,Tanabe:2015isb,Tanabe:2016opw}.\footnote{This membrane also forms a dual relativistic hydrodynamical description, providing a setting for understanding
the relation between black-hole dynamics and relativistic
hydrodynamics \cite{Halder:2026sdh,Halder:2026kwq}.}

In the charged membrane construction developed in ref.~\cite{Bhattacharyya:2015fdk}, the membrane velocity obeys the incompressibility condition at leading order,
$\widehat{\nabla}_{a}u^{a}=0$,
where $\widehat{\nabla}$ denotes the covariant derivative compatible with the membrane worldvolume metric induced from the auxiliary flat spacetime. A follow-up work in the neutral theory extends the membrane construction to one higher order in ref.~\cite{Dandekar:2016fvw}. Their analysis shows a non-negative divergence of the membrane velocity field, proportional to the quadratic shear term. The authors obtained this correction from a scalar Einstein constraint evaluated at the horizon. Later, in ref.~\cite{Bhattacharyya:2016nhn}, an entropy current was defined by pulling back the event-horizon area form to the membrane. At leading order, they showed that the membrane entropy current equals one quarter of the membrane velocity, which satisfies a non-negative divergence condition. Later on, ref.~\cite{Saha:2020zho} formulates the entropy current for general neutral large $D$ membranes, but the construction of a local entropy current and the explicit proof of non-negative entropy production for charged membranes remains unsolved.  
 
Building on this foundation, in this work, we investigate the local entropy production associated with the dynamics of charged black holes in the large $D$ membrane paradigm. We compute the subleading-order correction to the divergence of the membrane velocity for a charged black hole residing in a background flat spacetime. In the bulk black hole picture, the membrane velocity $u^a$ maps to the null generators of the physical event horizon in $D$ spacetime dimensions. Here, we evaluate the corresponding scalar constraint equation on the horizon via the null Raychaudhuri equation, which relates the evolution of the horizon expansion to the Ricci curvature along the generators. Pulling back these horizon quantities to the membrane, we determine the subleading, non-trivial contribution to the membrane expansion scalar. This calculation allows us to establish the non-negative divergence
of the membrane entropy current and obtain an explicit local entropy-production equation for charged membranes. Furthermore, this approach determines the first subleading correction to the membrane velocity divergence without requiring the construction of the complete bulk geometry at the next order in the large $D$ expansion.

The paper is organized as follows.  Section~\ref{sec:setup} reviews the
charged large $D$ setup and fixes conventions.  In
section~\ref{sec:scalarconstraint}, we explain the relation between the scalar Einstein constraint and the null Raychaudhuri equation and relate them to the membrane quantities.
Section~\ref{sec:maxwell} derives the Maxwell contribution. Section~\ref{sec:secondlaw} uses the Raychaudhuri equation to establish the positivity of the membrane expansion.
Section~\ref{sec:area} establishes the expression of the entropy current, which validates the local second law. The consistency check and regime of applicability are discussed in
section~\ref{sec:checks}, followed by a summary and future prospects in
section~\ref{sec:discussion}.
\section{Charged large $D$ membrane setup}
\label{sec:setup}

\subsection{Einstein--Maxwell Equation}

The Einstein--Maxwell action can be written as follows:
\begin{equation}
 S=
 \frac{1}{16\pi}
 \int \dd^{D}x\,
 \sqrt{-G}
 \left(
 R-\frac14 F_{MN}F^{MN}
 \right).
 \label{eq:action}
\end{equation}
where $G_{MN}$ denotes the full physical spacetime metric in $D$ spacetime dimension and
\begin{equation}
F_{MN}
=
\partial_M A_N-\partial_N A_M
\end{equation}
is the Maxwell field strength. 
Consequently, the Einstein-Maxwell field equation can be written as
\begin{equation}
 R_{MN}
 -
 \frac12 G_{MN}R
 =
 \frac12
 \left(
 F_{MP}F_{N}{}^{P}
 -
 \frac14G_{MN}F^{2}
 \right).
 \label{eq:einstein}
\end{equation}
Taking the trace of eq.~\eqref{eq:einstein} gives the expression of the Ricci scalar as
\begin{equation}
 R
 =
 \frac{D-4}{4(D-2)}F^{2},
\end{equation}
substituting the above value of Ricci scalar in eq.~\eqref{eq:einstein}, the Einstein-Maxwell field equation becomes
\begin{equation}
 R_{MN}
 =
 \frac12F_{MP}F_{N}{}^{P}
 -
 \frac{1}{4(D-2)}G_{MN}F^{2}.
 \label{eq:trace-reversed}
\end{equation}
\subsection{Auxiliary membrane geometry}
\label{subsec:auxiliary-geometry}

The large $D$ membrane is defined as the codimension-one hypersurface embedded in an auxiliary flat spacetime with metric $\eta_{MN}$, where the locus of a scalar function $\rho$ satisfies the condition $\rho=1$. We denote the covariant derivative compatible with the auxiliary flat metric $\eta_{MN}$ as  $\bar{\nabla}_{M}$.
The normal one-form to the constant-$\rho$ hypersurfaces is defined as
\begin{equation}
n_{M}
=
\frac{\partial_{M}\rho}{N_{\rho}},
\quad \text{with} \quad
N_{\rho}
=
\sqrt{
\eta^{MN}\partial_{M}\rho\,\partial_{N}\rho
}.
\label{eq:normal-definition}
\end{equation}
Indices on $n_{M}$ in the auxiliary geometry are raised and lowered using
$\eta_{MN}$. Therefore,
\begin{equation}
\eta^{MN}n_{M}n_{N}=1.
\label{eq:n-normalization}
\end{equation}
The normalized (with respect to the flat auxiliary metric) membrane velocity $u^{M}$ is defined as the tangent to the hypersurface. So it satisfies the following relations:
\begin{equation}
\eta_{MN}u^{M}u^{N}=-1,
\qquad
n_{M}u^{M}=0.
\label{eq:u-normalization}
\end{equation}
The second relation follows directly from the tangentiality condition
$u^{M}\partial_{M}\rho=0$. 
The induced metric on the timelike membrane is defined as
\begin{equation}
\gamma_{MN}
=
\eta_{MN}-n_{M}n_{N},
\label{eq:induced-metric}
\end{equation}
and the corresponding projector onto the membrane worldvolume is
\begin{equation}
\Pi_{M}{}^{N}
=
\delta_{M}{}^{N}-n_{M}n^{N}.
\label{eq:membrane-projector}
\end{equation}
The extrinsic curvature of the membrane can be written as
\begin{equation}
K_{MN}
=
\Pi_{M}{}^{P}
\Pi_{N}{}^{Q}
\bar{\nabla}_{P}n_{Q}.
\label{eq:extrinsic-curvature}
\end{equation}
We denote its trace by
\begin{equation}
K
=
\eta^{MN}K_{MN}.
\label{eq:trace-extrinsic-curvature}
\end{equation}
Hence, the trace of the extrinsic curvature can equivalently be written as
\begin{equation}
K=\bar{\nabla}_{M}n^{M}.
\label{eq:K-divergence-normal}
\end{equation}
We denote the covariant derivative compatible with the induced membrane worldvolume metric
$\gamma_{ab}$, by $\widehat{\nabla}_{a}$, satisfying
\begin{equation}
\widehat{\nabla}_{a}\gamma_{bc}=0.
\end{equation}
A tensor in the membrane worldvolume can be obtained by pulling an ambient tensor from the background flat spacetime. Similarly
$\widehat{\nabla}_{a}$ is obtained by projecting the auxiliary derivative
$\bar{\nabla}_{M}$ onto membrane indices\footnote{Capital Latin indices $M,N,\ldots$ denote indices in the $D$-dimensional ambient spacetime, whereas lower-case Latin indices $a,b,\ldots$ denote indices intrinsic to the $(D-1)$-dimensional membrane worldvolume.}. For example, for a tangent vector
$v^{b}$,
\begin{equation}
\widehat{\nabla}_{a}v^{b}
=
\Pi_{a}{}^{M}
\Pi^{b}{}_{N}
\bar{\nabla}_{M}v^{N}.
\label{eq:intrinsic-derivative}
\end{equation}
Accordingly, the membrane acceleration and the spatial projector
orthogonal to the velocity are defined as
\begin{equation}
a_{a}
=
u^{b}\widehat{\nabla}_{b}u_{a},
\qquad
P_{ab}
=
\gamma_{ab}+u_{a}u_{b}.
\label{eq:acceleration-projector}
\end{equation}
The membrane shear tensor is
\begin{equation}
\sigma_{ab}
=
P_{a}{}^{c}P_{b}{}^{d}
\widehat{\nabla}_{(c}u_{d)}
-
\frac{P_{ab}}{D-2}
\widehat{\nabla}_{c}u^{c},
\label{eq:shear-definition}
\end{equation}
where
\begin{equation}
X_{(ab)}
=
\frac12
\left(
X_{ab}+X_{ba}
\right).
\end{equation}
All quantities in
\eqref{eq:normal-definition}--\eqref{eq:shear-definition} are defined
using the auxiliary membrane geometry. They should therefore be
distinguished from quantities constructed using the full physical
black-hole metric $G_{MN}$ and its associated covariant derivative
$\nabla_{M}$, which will be used in later sections when discussing the null geometry of the event horizon.
\subsection{Charged large $D$ geometry}
\label{subsec:charged-geometry}

The leading large $D$ charged black-hole geometry is specified by the
membrane data $\rho$, $u_M$, and $Q$.  Introducing the null one-form as
\begin{equation}
O_M=n_M-u_M,
\label{eq:O-definition}
\end{equation}
With respect to the auxiliary metric we have,
\begin{equation}
O_M O^M=0,
\qquad
O_M n^M=1,
\qquad
O_M u^M=1.
\label{eq:O-properties}
\end{equation}
The leading metric and gauge field can be written as
\begin{align}
G^{(0)}_{MN}
&=
\eta_{MN}
+
F(\rho)\,O_M O_N,
\label{eq:leading-metric}
\\
A^{(0)}_M
&=
\sqrt{2}\,Q\,\rho^{-(D-3)}O_M ,
\label{eq:leading-gauge-field}
\end{align}
where
\begin{equation}
F(\rho)
=
(1+Q^2)\rho^{-(D-3)}
-
Q^2\rho^{-2(D-3)} .
\label{eq:F-rho}
\end{equation}
These expressions are the collective-coordinate form of the local
Reissner--Nordstr\"om geometry in the large $D$ expansion
\cite{Bhattacharyya:2015fdk}.
Here, we consider smooth membrane configurations with $K>0$ and
$K\sim\mathcal O(D)$, whose local charge profiles satisfy
\begin{equation}
1-Q(x)^2\geq\epsilon>0,
\label{eq:charge-domain}
\end{equation}
where $\epsilon$ is independent of $D$ throughout the
membrane region under consideration.
The membrane hypersurface $\rho=1$ is identified with the physical event
horizon.  In particular,
\begin{equation}
F(1)
=
(1+Q^2)-Q^2
=
1.
\label{eq:F-horizon}
\end{equation}
Consequently, on $\rho=1$,
\begin{align}
G^{(0)}_{MN}u^M u^N
&=
\eta_{MN}u^M u^N
+
F(1)(O_Mu^M)^2
\nonumber\\
&=
-1+1
=
0,
\label{eq:u-null-horizon}
\end{align}
so that $u^M$, which is unit timelike with respect to the auxiliary metric
$\eta_{MN}$, becomes null with respect to the physical leading metric
$G^{(0)}_{MN}$ on the horizon.  Moreover,
\begin{equation}
G^{(0)}_{MN}u^N
=
u_M+F(1)O_M(O_Nu^N)
=
u_M+O_M
=
n_M,
\label{eq:Gu-n}
\end{equation}
or equivalently,
\begin{equation}
G_{(0)}^{MN}n_N=u^M .
\label{eq:Ginv-n-u}
\end{equation}
The first subleading corrections to the metric and gauge field can be
decomposed into scalar, vector and tensor sectors as
\begin{align}
G_{MN}
={}&
\eta_{MN}
+
F(\rho)O_MO_N
+
H^{(S)}O_MO_N
+
2O_{(M}H^{(V)}_{N)}
+
H^{(T)}_{MN}
+
H^{(\mathrm{Tr})}\mathcal{P}_{MN}
+\mathcal{O}(D^{-2}),
\label{eq:corrected-metric}
\\
A_M
={}&
\sqrt{2}\,Q\,\rho^{-(D-3)}O_M
+
A^{(S)}O_M
+
A^{(V)}_M
+
\mathcal{O}(D^{-2}),
\label{eq:corrected-gauge-field}
\end{align}
where
\begin{equation}
\mathcal{P}_{MN}
=
\eta_{MN}
-
n_M n_N
+
u_M u_N,
\label{eq:ambient-screen-projector}
\end{equation}
which is orthogonal to both $n^M$ and $u^M$ in the
auxiliary geometry.  The vector corrections $H^{(V)}_M$ and
$A^{(V)}_M$ are transverse to $n^M$ and $u^M$, while $H^{(T)}_{MN}$ is also a transverse and traceless tensor. The large $D$ scalings of those corrections are listed in ref.~\cite{Bhattacharyya:2015fdk}. They are scaled as follows:
\begin{align}
&A^{(S)}
=
\mathcal{O}\left(D^{-1}\right),
\qquad
A^{(V)}_M
=
\mathcal{O}\left(D^{-1}\right),
\qquad
H^{(S)}
=
\mathcal{O}\left(D^{-1}\right),
\notag\\ &
H^{(V)}_M
=
\mathcal{O}\left(D^{-1}\right),
\qquad
H^{(T)}_{MN}
=
\mathcal{O}\left(D^{-1}\right),
\quad H^{(\mathrm{Tr})}
=
\mathcal{O}\left(D^{-3}\right).
\label{eq:first-order-scaling}
\end{align}
Thus, all non-trace corrections to the metric and gauge field arise earlier in the perturbation series at $1/D$, with the suppressed trace correction.
We fix the membrane fields order by order using the gauge and
horizon normalization conditions. Let
\[
\Delta G_{MN}= G_{MN}-G^{(0)}_{MN},
\qquad
\Delta A_M= A_M-A^{(0)}_M.
\]
The gauge conditions are
\[
\Delta G_{MN}O^N=0,
\qquad
\Delta A_M O^M=0,
\]
and the horizon normalization conditions are
\[
\left.\Delta G_{MN}n^N\right|_{\rho=1}=0,
\qquad
\left.\Delta A_M n^M\right|_{\rho=1}=0.
\]
Indices on $O_M$, $n_M$, and $u_M$ in these conditions are
raised with $\eta^{MN}$. Since $u^N=n^N-O^N$, we obtain
$\Delta G_{MN}u^N=0$ on the horizon. Consequently, the full
metric satisfies
\begin{equation}
G_{MN}u^N
=
n_M.
\end{equation}
Following this we have
\begin{equation}
G_{MN}u^M u^N
=
n_N u^N=0
\end{equation}
Thus, $u^M$ is the null generator of the $\rho=1$ hypersurface in the
physical black hole geometry.

\section{The horizon scalar constraint and null Raychaudhuri equation}
\label{sec:scalarconstraint}
\subsection{Horizon scalar constraint in the neutral and charged theories}
\label{subsec:scalar-constraint}

In the first subleading analysis of the neutral large $D$ membrane,
ref.~\cite{Dandekar:2016fvw} determines the first non-vanishing
contribution to the velocity divergence from a scalar gravitational
constraint equation evaluated at the horizon.  It is useful to relate that construction directly to the charged Einstein--Maxwell system considered here.
For the neutral theory, let
\begin{equation}
\mathcal{E}^{(0)}_{MN}
=
R_{MN}
-\frac{1}{2}R\,G_{MN}
\end{equation}
The Einstein equation is
$\mathcal{E}^{(0)}_{MN}=0$.  The gravitational constraint associated with the foliation by constant-$\rho$ hypersurfaces can be written as
\begin{equation}
\mathcal{C}_{M}
=
    \mathcal{E}^{(0)}_{MN}G^{NP}n_{P}=0.
\label{eq:constraint-vector}
\end{equation}
Contracting this equation with $u^M$ at the event horizon produces the
corresponding scalar constraint.
In the charged theory, the same construction applies, but now
$\mathcal{E}_{MN}$ represents for the full Einstein--Maxwell equation (eq.~\eqref{eq:trace-reversed}). So the equation is given as
\begin{equation}
\mathcal{E}_{MN}
=
R_{MN}
-
 \frac12F_{MP}F_{N}{}^{P}
 +
 \frac{1}{4(D-2)}G_{MN}F^{2}
=0.
\label{eq:einstein-maxwell-equation}
\end{equation}
As established in the preceding subsection, at the horizon we can write
\begin{equation}
G^{MN}n_{N}=u^{M}.
\label{eq:horizon-normalization-use}
\end{equation}
Therefore, for a charged configuration, the gravitational scalar constraint equation at the event horizon becomes
\begin{equation}
u^{M}\mathcal{C}_{M}
=
\mathcal{E}_{MN}u^{M}u^{N}
=
0.
\label{eq:scalar-constraint-uu}
\end{equation}
Since $u^{M}$ is null with respect to the physical metric on the horizon,
$G_{MN}u^{M}u^{N}=0$, all terms in eq.~\eqref{eq:einstein-maxwell-equation} proportional to $G_{MN}$ drop out
of this contraction.  The corresponding charged scalar gravitational constraint thus
reduces to
\begin{equation}\label{Constraint}
R_{MN}u^{M}u^{N}
=
\frac{1}{2}
F_{MP}u^{M}
F_{N}{}^{P}u^{N}.
\end{equation}
Introducing
\begin{equation}
R_{uu}= R_{MN}u^{M}u^{N},
\quad \text{and} \quad
F_{uP}= u^{M}F_{MP},
\end{equation}
to eq.~\eqref{Constraint}, we can write
\begin{equation}
R_{uu}
=
\frac{1}{2}F_{uP}F_{u}{}^{P}.
\label{eq:Ruu-Maxwell}
\end{equation}
In the neutral limit $F_{MN}=0$, this reduces to $R_{uu}=0$, which is
the corresponding horizon scalar Einstein constraint.
\subsection{Non-affinity of the horizon generator}
\label{subsec:horizon-nonaffinity}
As established above, the physical event horizon is the hypersurface at $\rho=1$, with the normal one-form
\begin{equation}
n_M
=
\frac{\partial_M\rho}{N_\rho}
=
\frac{\nabla_M\rho}{N_\rho},
\label{eq:n-gradient-physical}
\end{equation}
Since $\rho=1$ is a null hypersurface of the physical metric, its normal is
null on the horizon,
\begin{equation}
G^{MN}n_Mn_N=0
\label{eq:n-null-physical}
\end{equation}
as eq.~\eqref{eq:horizon-normalization-use} identifies its
metric-raised normal with the horizon generator $u^M$.

We now verify directly, that this
generator satisfies the non-affinely parametrized null geodesic
equation.  Since the equality
$u^N=G^{MN}n_M$ holds everywhere on the horizon and $u^N$ is tangent
to the horizon, differentiation along $u^N$ on the horizon gives
\begin{equation}
u^N\nabla_Nu^M
=
G^{MP}u^N\nabla_Nn_P ,
\label{eq:u-derivative-n}
\end{equation}
where the metric compatibility relation $\nabla_MG^{NP}=0$, is used.

Using $n_M=N_\rho^{-1}\nabla_M\rho$, we obtain
\begin{align}
u^N\nabla_Nn_M
&=
-\left(u^N\nabla_N\ln N_\rho\right)n_M
+
\frac{1}{N_\rho}
u^N\nabla_N\nabla_M\rho
\nonumber\\
&=
-\left(u^N\nabla_N\ln N_\rho\right)n_M
+
\frac{1}{N_\rho}
u^N\nabla_M\nabla_N\rho ,
\label{eq:unabla-n-step1}
\end{align}
where the covariant derivatives commute on the scalar $\rho$.
On the horizon, using
$u^N=G^{NQ}n_Q$ and $\nabla_N\rho=N_\rho n_N$, the second term becomes
\begin{align}
\frac{1}{N_\rho}
u^N\nabla_M\nabla_N\rho
&=
\frac{1}{N_\rho}
G^{NQ}n_Q
\nabla_M\left(N_\rho n_N\right)
\nonumber\\
&=
G^{NQ}n_Qn_N\,\nabla_M\ln N_\rho
+
\frac12
\nabla_M\left(G^{NQ}n_Nn_Q\right).
\label{eq:unabla-n-step2}
\end{align}
The first term on the right-hand side vanishes on $\rho=1$ by
eq.~\eqref{eq:n-null-physical}.  Hence
\begin{equation}
u^N\nabla_Nn_M
=
\frac12
\nabla_M\left(G^{PQ}n_Pn_Q\right)
-
\left(u^N\nabla_N\ln N_\rho\right)n_M.
\label{eq:unabla-n}
\end{equation}
The scalar $G^{PQ}n_Pn_Q$ vanishes identically on the hypersurface
$\rho=1$. Consequently, its gradient on the horizon is
normal to the hypersurface.  Since $n_M$ is the normal one-form, we can write
\begin{equation}
\left.
\nabla_M\left(G^{PQ}n_Pn_Q\right)
\right|_{\rho=1}
\propto n_M .
\label{eq:gradient-norm-normal}
\end{equation}
Eq.~\eqref{eq:unabla-n} therefore shows that
$u^N\nabla_Nn_M$ is proportional to $n_M$.  Raising the index with the
physical metric and using $G^{MN}n_N=u^M$ gives
\begin{equation}
u^N\nabla_Nu^M
=
\kappa u^M
\label{eq:nonaffine}
\end{equation}
This is the non-affinely parametrized null geodesic equation for the
horizon generators.
The leading large $D$ value of $\kappa$ can be extracted from the same
relation.  The inverse of the leading metric
\eqref{eq:leading-metric} can be written as
\begin{equation}
G_{(0)}^{MN}
=
\eta^{MN}
-
F(\rho)O^MO^N .
\label{eq:leading-inverse-metric}
\end{equation}
Using $\eta^{MN}n_Mn_N=1$, and $O^Mn_M=1$ it follows
\begin{equation}
G_{(0)}^{MN}n_Mn_N
=
1-F(\rho).
\label{eq:norm-normal-leading}
\end{equation}
Since $F(\rho,Q)|_{(\rho=1)}=1$ for arbitrary $Q$,
\begin{equation}
\left.
\partial_QF
\right|_{\rho=1}
=
0 ,
\end{equation}
and therefore the leading-order part of the gradient in
eq.~\eqref{eq:unabla-n} is purely normal. So,
\begin{equation}
\left.
\frac12
\nabla_M\left(G^{PN}n_Pn_N\right)
\right|_{\rho=1}
=
\frac{N_\rho}{2}
\left.
\partial_\rho(1-F)
\right|_{\rho=1}
n_M
+
\mathcal{O}(1).
\label{eq:norm-gradient-horizon}
\end{equation}
The first-subleading corrections to the metric can contribute only at
$\mathcal{O}(1)$ to this expression and hence do not modify its leading
order ($\mathcal{O}(D)$) part. Using eq.~\eqref{eq:unabla-n} we have
\begin{equation}
\kappa
=
\frac{N_\rho}{2}
\left.
\partial_\rho(1-F)
\right|_{\rho=1}
-
u^N\nabla_N\ln N_\rho
+
\mathcal{O}(1).
\label{eq:kappa-intermediate}
\end{equation}
Again using the large D scaling we have $u^N\nabla_N\ln N_\rho=\mathcal{O}(1)$, whereas the first term is
$\mathcal{O}(D)$. Thus, only the first term is required for the leading-order expression of $\kappa$. As
\begin{equation}
F(\rho)
=
(1+Q^2)\rho^{-(D-3)}
-
Q^2\rho^{-2(D-3)},
\end{equation}
we have
\begin{equation}
\left.
\partial_\rho(1-F)
\right|_{\rho=1}
=
(D-3)(1-Q^2).
\label{eq:F-radial-derivative}
\end{equation}
The subsidiary condition of the charged large $D$ construction
\cite{Bhattacharyya:2015fdk} can be written as,
\begin{equation}
\rho\,\bar{\nabla}^{2}\rho
=
(D-2)\,
\bar{\nabla}_M\rho\,\bar{\nabla}^{M}\rho ,
\label{eq:rho-subsidiary}
\end{equation}
together with
$N_\rho^2=\bar{\nabla}_M\rho\,\bar{\nabla}^{M}\rho$, gives at
$\rho=1$
\begin{align}
K
=
\bar{\nabla}_Mn^M
&=
\frac{\bar{\nabla}^{2}\rho}{N_\rho}
-
\frac{n^M\bar{\nabla}_MN_\rho}{N_\rho}
\nonumber\\
&=
(D-2)N_\rho
-
\frac{n^M\bar{\nabla}_MN_\rho}{N_\rho}.
\label{eq:K-N-relation-exact}
\end{align}
Since the second term is $\mathcal{O}(1)$ while $K=\mathcal{O}(D)$,
\begin{equation}
N_\rho
=
\frac{K}{D-2}
+
\mathcal{O}(D^{-1}).
\label{eq:N-K-relation}
\end{equation}
Combining eqs.~\eqref{eq:kappa-intermediate},
\eqref{eq:F-radial-derivative}, and
\eqref{eq:N-K-relation}, we obtain
\begin{equation}
\kappa
=
\frac{D-3}{2(D-2)}
K(1-Q^2)
+
\mathcal{O}(1),
\end{equation}
and hence
\begin{equation}
\kappa
=
\frac{K}{2}(1-Q^2)
+
\mathcal{O}(1)
\label{eq:kappa-leading}
\end{equation}
For the charge profiles satisfying
eq.~\eqref{eq:charge-domain}, this gives
$\kappa=\mathcal O(D)$.
\subsection{Null Raychaudhuri equation}
\label{subsec:null-Raychaudhuri}
The non-affine null Raychaudhuri equation in $D$ spacetime dimensions can be written using ref~
\cite{Raychaudhuri:1955,Sachs:1961,Amsel:2007mh} as
\begin{equation}
u^M\nabla_M\theta_{\rm H}
=
\kappa\theta_{\rm H}
-
\frac{\theta_{\rm H}^{2}}{D-2}
-
\sigma_{\rm H}^{2}
+
\omega_{\rm H}^{2}
-
R_{MN}u^Mu^N .
\label{eq:ray-general}
\end{equation}
Here $\theta_H$, $\sigma_H$, and $\omega_H$ denote the expansion,
shear, and twist of the physical null horizon congruence.
On the horizon,
$G_{MN}u^N=n_M=N_\rho^{-1}\nabla_M\rho$.
The generators are therefore hypersurface orthogonal, and their twist vanishes by Frobenius' theorem,
$\omega_H=0$ see refs.~\cite{Raychaudhuri:1955,Sachs:1961}. Hence eq.~\eqref{eq:ray-general} reduces to the following form
\begin{equation}
R_{uu}
=
\kappa\theta_{\rm H}
-
u^M\nabla_M\theta_{\rm H}
-
\sigma_{\rm H}^{2}
-
\frac{\theta_{\rm H}^{2}}{D-2},
\label{eq:ray}
\end{equation}
where
\begin{equation}
R_{uu}= R_{MN}u^Mu^N .
\end{equation}
The above analysis shows that eq.~\eqref{eq:ray} constrains the scalar u-u projection of the Ricci tensor.  
\subsection{Relation between the horizon and membrane kinematics}
\label{subsec:horizon-membrane-kinematics}

We now relate the expansion and shear of the physical null horizon to the corresponding quantities defined intrinsically on the auxiliary timelike membrane.
At a point on $\rho=1$, choose a basis $e_i^{\,M}$,
$i=1,\ldots,D-2$, spanning the spatial directions tangent to the
membrane and orthogonal to its velocity.  Thus, with all contractions
in the following two conditions performed using the auxiliary flat metric,
\begin{equation}
n_M e_i^{\,M}=0,
\qquad
u_M e_i^{\,M}=0 .
\label{eq:spatial-basis}
\end{equation}
Since $O_M=n_M-u_M$, it follows immediately that
\begin{equation}
O_Me_i^{\,M}=0 .
\label{eq:O-screen-zero}
\end{equation}
The spatial metric induced from the auxiliary membrane geometry on this
$(D-2)$ dimensional subspace is
\begin{equation}
h_{ij}
=
\eta_{MN}e_i^{\,M}e_j^{\,N}.
\label{eq:membrane-spatial-metric}
\end{equation}
Its determinant will be denoted by
\begin{equation}
h=\det(h_{ij}).
\end{equation}
The corresponding physical horizon metric is the pullback of the full
spacetime metric,
\begin{equation}
h^{\rm H}_{ij}
=
G_{MN}e_i^{\,M}e_j^{\,N},
\qquad
h_{\rm H}=\det(h^{\rm H}_{ij}).
\label{eq:horizon-spatial-metric}
\end{equation}
At leading order we have
\begin{equation}
G^{(0)}_{MN}
=
\eta_{MN}+F(\rho)O_MO_N .
\end{equation}
Using eq.~\eqref{eq:O-screen-zero}, its pullback to the spatial subspace gives
\begin{equation}
h^{\rm H(0)}_{ij}
=
h_{ij}.
\label{eq:leading-spatial-metrics}
\end{equation}
Thus, the physical and auxiliary spatial metrics agree at leading
order, their difference begins with the subleading metric corrections. This difference can be written as
\begin{equation}
\delta h_{ij}
=
h^{\rm H}_{ij}-h_{ij}.
\label{eq:delta-h}
\end{equation}
The scalar and vector metric corrections are proportional to $O_M$ and will have
no contribution in $\delta h_{ij}$ because of eq.~\eqref{eq:O-screen-zero}.  Therefore, the
correction to the spatial metric is supplied by the $(D-2)$ dimensional spatial part of the tensor and trace sectors of the metric in eq.~\eqref{eq:corrected-metric}. So we have
\begin{equation}
\delta h_{ij}
=
H^{(T)}_{ij}
+
H^{(\mathrm{Tr})}h_{ij}
+\mathcal{O}(D^{-2}) ,
\label{eq:screen-metric-correction}
\end{equation}
where
\begin{equation}
H^{(T)}_{ij}
=
H^{(T)}_{MN}e_i^{\,M}e_j^{\,N}.
\end{equation}
From eq.~\eqref{eq:first-order-scaling} it follows
\begin{equation}
H^{(T)}_{ij}=\mathcal{O}(D^{-1}),
\qquad
h^{ij}H^{(T)}_{ij}=0,
\qquad
H^{(\mathrm{Tr})}=\mathcal{O}(D^{-3}).
\label{eq:screen-correction-orders}
\end{equation}
It is important to know that the omitted $\mathcal{O}(D^{-2})$ terms in eq.~\eqref{eq:screen-metric-correction} are the contribution from the pure traceless tensor sector, as the full trace coefficient is $\mathcal{O}(D^{-3})$~\cite{Bhattacharyya:2015fdk}. Now we define the ratio of the determinant of the induced spatial metrics as follows
\begin{equation}
\zeta
=
\frac{\sqrt{h_{\rm H}}}{\sqrt{h}} .
\label{eq:zeta-definition}
\end{equation}
Using
\begin{equation}
h^{\rm H}_{ij}=h_{ij}+\delta h_{ij},
\end{equation}
the determinant expansion gives
\begin{align}
\ln\zeta
&=
\frac12
\operatorname{Tr}\ln
\left(
\delta^i{}_j+h^{ik}\delta h_{kj}
\right)
\nonumber\\
&=
\frac12 h^{ij}\delta h_{ij}
-
\frac14
h^{ik}\delta h_{kj}
h^{jl}\delta h_{li}
+\cdots .
\label{eq:zeta-determinant-expansion}
\end{align}
As per eqs.~\eqref{eq:screen-metric-correction}~and~\eqref{eq:screen-correction-orders}, the contribution from the first term in the above equation produces only
$\mathcal{O}(D^{-2})$ terms. The second term, as well as the other contributions, also contributes to $\mathcal{O}(D^{-2})$ by
eq.~\eqref{eq:screen-correction-orders}, as the spacetime obeys $SO(D-p-2)$ isometry \cite{Bhattacharyya:2015fdk} (See appendix~\ref{Some} for details). So we have
\begin{equation}
\ln\zeta=\mathcal{O}(D^{-2}),
\qquad
\zeta=1+\mathcal{O}(D^{-2}) .
\label{eq:zeta-order}
\end{equation}
Now we compare the two expansions. The expansion of the physical null horizon generators is defined by the rate of change of the horizon area element, leading to the following expression:
\begin{equation}
\theta_{\rm H}
=
\frac{1}{\sqrt{h_{\rm H}}}
\mathcal{L}_u\sqrt{h_{\rm H}} .
\label{eq:theta-H-area}
\end{equation}
The corresponding expansion scalar for the membrane velocity in the auxiliary geometry can be written as
\begin{equation}
\widehat{\nabla}_a u^a
=
\frac{1}{\sqrt{h}}
\mathcal{L}_u\sqrt{h}.
\label{eq:membrane-area-expansion}
\end{equation}
Here $\mathcal{L}_u$ represents the Lie derivative along u. Using eqs.~\eqref{eq:zeta-definition}~\eqref{eq:theta-H-area}~and~\eqref{eq:membrane-area-expansion} we get
\begin{align}
\theta_{\rm H}
&=
\frac{1}{\zeta\sqrt{h}}
\mathcal{L}_u
\left(
\zeta\sqrt{h}
\right)
\nonumber\\
&=
\widehat{\nabla}_a u^a
+
u^a\widehat{\nabla}_a\ln\zeta .
\label{eq:thetaH-divu-exact}
\end{align}
Using eq.~\eqref{eq:zeta-order} in the above expression yields
\begin{equation}
\theta_{\rm H}
=
\widehat{\nabla}_a u^a
+
\mathcal{O}(D^{-2}) .
\label{eq:thetaH-divu}
\end{equation}
Now, to find the relation between the physical horizon shear and the membrane shear, define the symmetric deformation
tensor of the physical horizon spatial metric by
\begin{equation}
B^{\rm H}_{ij}=\frac12\mathcal{L}_u h^{\rm H}_{ij},
\label{eq:BH-definition}
\end{equation}
so that
\begin{equation}
\theta_{\rm H}
=
(h_{\rm H})^{ij}B^{\rm H}_{ij}.
\end{equation}
The physical horizon shear is its traceless part,
\begin{equation}
\sigma^{\rm H}_{ij}
=
B^{\rm H}_{ij}
-
\frac{\theta_{\rm H}}{D-2}
h^{\rm H}_{ij},
\quad \text{with} \quad
\sigma_{\rm H}^{2}
=
(h_{\rm H})^{ik}(h_{\rm H})^{jl}
\sigma^{\rm H}_{ij}\sigma^{\rm H}_{kl}.
\label{eq:horizon-shear-definition}
\end{equation}
Similarly, the membrane deformation tensor can be written as
\begin{equation}
B_{ij}=\frac12\mathcal{L}_u h_{ij}.
\label{eq:B-definition}
\end{equation}
Consequently, the membrane shear\footnote{Which can be equivalently written as the pullback of the membrane shear (in eq.~\eqref{eq:shear-definition}) to the same spatial basis as
$\sigma_{ij} = e_i^{\,a}e_j^{\,b}\sigma_{ab}.$} becomes 
\begin{equation}
\sigma_{ij}
=
B_{ij}
-
\frac{\widehat{\nabla} \cdot u}{D-2}
h_{ij},
\quad \text{with} \quad
\sigma^{2}
=
h^{ik}h^{jl}
\sigma_{ij}\sigma_{kl}.
\label{eq:membrane-shear-definition}
\end{equation}
Since $h^{\rm H}_{ij}=h_{ij}+\mathcal{O}(D^{-1})$, using eqs.~\eqref{eq:BH-definition}~and~\eqref{eq:B-definition} we have

\begin{equation}
B^{\rm H}_{ij}
=
B_{ij}
+
\mathcal{O}(D^{-1}),
\label{eq:BH-B-order}
\end{equation}
Also using eq.~\eqref{eq:thetaH-divu}, we obtain
\begin{equation}
\sigma^{\rm H}_{ij}
=
\sigma_{ij}
+
\mathcal{O}(D^{-1}).
\label{eq:shear-tensor-matching}
\end{equation}
The inverse spatial metrics similarly satisfy
\begin{equation}
(h_{\rm H})^{ij}
=
h^{ij}
+
\mathcal{O}(D^{-1}).
\end{equation}
Therefore, on contracting the shear tensors we have
\begin{equation}
\sigma_{\rm H}^{2}
=
\sigma_{ab}\sigma^{ab}
+
\mathcal{O}(D^{-1}) .
\label{eq:shear-matching}
\end{equation}

\subsection{Large $D$ scaling of the scalar equation}
\label{subsec:large-D-Raychaudhuri}

From refs.~\cite{Bhattacharyya:2016nhn,Bhattacharyya:2015fdk} we have
$\widehat{\nabla}_a u^a\sim\mathcal{O}(1/D)$. So eq.~\eqref{eq:thetaH-divu} implies
\begin{equation}
\theta_{\rm H}
\sim
\mathcal{O}(D^{-1}).
\label{eq:theta-order}
\end{equation}
Together with eq.~\eqref{eq:kappa-leading}, we have
\begin{equation}
\kappa\sim \mathcal{O}(D),
\qquad
\kappa\theta_{\rm H}
\sim
\mathcal{O}(1).
\label{eq:kappa-theta-order}
\end{equation}
Consequently,
\begin{equation}
u^M\nabla_M\theta_{\rm H}
\sim
\mathcal{O}(D^{-1}),
\label{eq:theta-dot-order}
\end{equation}
and
\begin{equation}
\frac{\theta_{\rm H}^{2}}{D-2}
\sim
\mathcal{O}(D^{-3}).
\label{eq:theta-square-order}
\end{equation}
The horizon shear satisfies
\begin{equation}
\sigma_{\rm H}^{2}
\sim\sigma^{2}\sim
\mathcal{O}(1).
\label{eq:sigma-H-order}
\end{equation}
Keeping the order-$D^{0}$ terms in
eq.~\eqref{eq:ray} at leading order we get
\begin{equation}
R_{uu}
=
\kappa\theta_{\rm H}
-
\sigma_{\rm H}^{2}
+
\mathcal{O}(D^{-1}).
\label{eq:ray-leading}
\end{equation}
Using eqs.\eqref{eq:kappa-leading},
\eqref{eq:thetaH-divu}, and \eqref{eq:shear-matching}, the above equation becomes
\begin{equation}
R_{uu}
=
\frac{K}{2}(1-Q^2)\,
\widehat{\nabla}_a u^a
-
\sigma_{ab}\sigma^{ab}
+
\mathcal{O}(D^{-1}).
\label{eq:ray-membrane-variables}
\end{equation}

\section{Scalar projection of the Einstein--Maxwell equation}
\label{sec:maxwell}

In this section, we evaluate the Maxwell term in the horizon scalar constraint.
We use $\bar\nabla_M$ for the auxiliary connection and
$\widehat\nabla_a$ for the induced membrane connection, as defined in section~\ref{subsec:auxiliary-geometry}. Indices on auxiliary geometric quantities are raised
with $\eta^{MN}$, and intrinsic indices with $\gamma^{ab}$.
The contraction $F_{uM}F_u{}^M$, however, uses the full inverse
metric $G^{MN}$. All horizon identities below are evaluated
at $\rho=1$.
Since the Christoffel connection is torsion-free, we can write
\begin{equation}
F_{MN}
=2\partial_{[M}A_{N]}
=2\bar\nabla_{[M}A_{N]}.
\label{eq:F-definition}
\end{equation}
Where the square bracket implies the halved anti-symmetrization. We first determine the part of $F_{uM}$ tangent to the spatial
directions of the membrane. With $\mathcal P_M{}^N$ denoting the
projector orthogonal to both $u^M$ and $n^M$, consider
\begin{equation}
\mathcal P_M{}^N F_{uN}
=\mathcal P_M{}^N u^P F_{PN}.
\label{eq:screen-Fu}
\end{equation}
For the leading gauge field
\begin{equation}
A_M^{(0)}
=\sqrt{2}\,Q\,\rho^{-(D-3)}O_M,
\end{equation}
we have
\begin{align}
F_{MN}^{(0)}
={}&
\sqrt{2}\rho^{-(D-3)}
\left[
(\bar\nabla_M Q)O_N-(\bar\nabla_N Q)O_M
\right]
\nonumber\\
&+\sqrt{2}Q\rho^{-(D-3)}
\left(
\bar\nabla_M O_N-\bar\nabla_N O_M
\right)
\nonumber\\
&-\sqrt{2}(D-3)Q\rho^{-(D-2)}
\left[
(\bar\nabla_M\rho)O_N-(\bar\nabla_N\rho)O_M
\right].
\label{eq:F-leading}
\end{align}
Since $n_M=N_\rho^{-1}\bar\nabla_M\rho$ and the covariant
derivatives commute on the scalar $\rho$, we have the following relation:
\[
\bar\nabla_P n_N-\bar\nabla_N n_P
=n_P\bar\nabla_N\ln N_\rho
-n_N\bar\nabla_P\ln N_\rho.
\]
Contracting with $\mathcal P_M{}^N u^P$ and using
$u^Pn_P=0$ and $\mathcal P_M{}^Nn_N=0$ therefore gives
\begin{equation}
\mathcal P_M{}^N u^P
\left(
\bar\nabla_P n_N-\bar\nabla_N n_P
\right)=0.
\end{equation}
Moreover, the normalization $u^Pu_P=-1$ implies
\[
\mathcal P_M{}^N u^P\bar\nabla_N u_P
=\frac12\mathcal P_M{}^N\bar\nabla_N(u^Pu_P)=0.
\]
Using $O_M=n_M-u_M$, it follows that
\begin{equation}
\mathcal P_M{}^N u^P
\left(
\bar\nabla_P O_N-\bar\nabla_N O_P
\right)=-a_M,
\end{equation}
where
\begin{equation}
a_M
=
\mathcal P_M{}^N u^P\bar\nabla_P u_N.
\label{eq:ambient-acceleration}
\end{equation}
This is the ambient representation of the intrinsic membrane
acceleration. To show this, we use the pullback $\Pi_a{}^M$ introduced in section~\ref{subsec:auxiliary-geometry}. We have
\[
\widehat\nabla_bu_a
=\Pi_b{}^P\Pi_a{}^N\bar\nabla_Pu_N,
\qquad
\Pi_a{}^M\mathcal P_M{}^N=P_a{}^c\Pi_c{}^N.
\]
Using $u^P=\Pi_b{}^P u^b$, we obtain
\begin{equation}\label{aPullback}
    \Pi_d{}^M a_M
=P_d{}^c u^b\widehat\nabla_bu_c
=u^b\widehat\nabla_bu_d
=a_d.
\end{equation}
Now using eqs.\eqref{eq:F-leading}~\eqref{eq:screen-Fu}~and~\eqref{eq:O-properties} at $\rho=1$ we get
\begin{equation}
\mathcal P_M{}^N F^{(0)}_{uN}
=-\sqrt{2}\,V_M,
\label{eq:leading-screen-F}
\end{equation}
where
\begin{equation}
V_M
=
\mathcal P_M{}^N\bar\nabla_NQ+Qa_M.
\label{eq:V-ambient}
\end{equation}
Using $\Pi_b{}^N\bar\nabla_NQ=\widehat\nabla_bQ$ for the scalar 
$Q$, and eq.~\eqref{aPullback}, the pullback of $V_M$ to the membrane is
\begin{equation}
V_d
=\Pi_d{}^M V_M
=P_d{}^b\widehat\nabla_bQ+Qa_d.
\label{eq:V-def}
\end{equation}
We next check the contribution of the first subleading gauge field
correction to eq.~\eqref{eq:leading-screen-F}. The charged large $D$
solution of ref.~\cite{Bhattacharyya:2015fdk} is
\begin{equation}
A_M
=\sqrt{2}\,Q\rho^{-(D-3)}O_M
+A^{(S)}O_M+A_M^{(V)}
+\mathcal O(D^{-2}),
\label{eq:A-corrected}
\end{equation}
with
\begin{align}
A^{(S)}
={}&
\frac{1}{D}
\Bigg[
\sqrt{2}\,Q\,D(\rho-1)\rho^{-D}\,
\mathfrak S_{(1)}
\nonumber\\
&\hspace{18mm}
+2\sqrt{2}\,
\frac{Q^3}{1-Q^2}\,
\rho^{-D}\Upsilon_A(\rho)\,
\mathfrak S_{(2)}
\Bigg]
+\mathcal O(D^{-2}),
\label{eq:AS-explicit}
\\[2mm]
A_M^{(V)}
={}&
-\frac{\sqrt{2}}{D}\,Q\rho^{-D}
\Bigg[
D(\rho-1)
\left(
\mathfrak V_M^{(1)}-Q^2\mathfrak V_M^{(2)}
\right)
\nonumber\\
&\hspace{18mm}
-Q^2
\left\{
1+\log\!\left(1-Q^2\rho^{-D}\right)
\right\}
\mathfrak V_M^{(2)}
\Bigg]
+\mathcal O(D^{-2}),
\label{eq:AV-explicit}
\end{align}
where
\begin{equation}
\Upsilon_A(\rho)
=\int_0^{D(\rho-1)}
dx\,\log\!\left(1-Q^2e^{-x}\right).
\label{eq:UpsilonA}
\end{equation}
The scalar and vector data appearing here are
\begin{align}
\mathfrak S_{(1)}
&=\frac{D}{K}
\left[
\frac{u\cdot\bar\nabla Q}{Q}
-u^MK_{MN}u^N
+\frac{u\cdot\bar\nabla K}{K}
\right],
\label{eq:S1-data}
\\
\mathfrak S_{(2)}
&=\frac{D}{K}
\left[
u^MK_{MN}u^N
-\frac{u\cdot\bar\nabla K}{K}
\right],
\label{eq:S2-data}
\\
\mathfrak V_M^{(1)}
&=\frac{D}{K}\,
\mathcal P_M{}^N
\left[
\frac{\bar\nabla_N K}{K}
+u^P\bar\nabla_P u_N
\right],
\label{eq:V1-data}
\\
\mathfrak V_M^{(2)}
&=\frac{D}{K}\,
\mathcal P_M{}^N
\left[
\frac{\bar\nabla_N K}{K}
-u^P\bar\nabla_P u_N
\right].
\label{eq:V2-data}
\end{align}
The explicit factor $1/D$ in
eqs.~\eqref{eq:AS-explicit}--\eqref{eq:AV-explicit}
is not by itself sufficient to neglect these terms, since a radial
derivative can produce a factor of $D$. We therefore examine their
contribution to eq.\eqref{eq:screen-Fu} directly.
For the scalar correction $A^{(S)}$ we have the following constraint:
\begin{equation}
A^{(S)}\big|_{\rho=1}=0.
\label{eq:AS-zero}
\end{equation}
The displayed terms in eq~\eqref{eq:AS-explicit} satisfy this
condition because both $D(\rho-1)$ and $\Upsilon_A(\rho)$
vanish there. Its radial derivative can nevertheless be
of order unity. We denote the scalar subleading correction to the gauge field as
\begin{equation}
\delta A_M^{(S)}=A^{(S)}O_M,
\end{equation}
now we obtain
\begin{align}
\delta F_{MN}^{(S)}
={}&
(\bar\nabla_M A^{(S)})O_N
-(\bar\nabla_N A^{(S)})O_M
\nonumber\\
&+A^{(S)}
\left(
\bar\nabla_M O_N-\bar\nabla_N O_M
\right).
\label{eq:deltaF-scalar}
\end{align}
The following contraction gives
\begin{align}
\mathcal P_M{}^N u^P\delta F_{PN}^{(S)}
={}&
-\mathcal P_M{}^N\bar\nabla_N A^{(S)}
\nonumber\\
&+A^{(S)}
\mathcal P_M{}^N u^P
\left(
\bar\nabla_P O_N-\bar\nabla_N O_P
\right),
\label{eq:scalar-screen-direct}
\end{align}
where we have used $u^PO_P=1$ and $\mathcal P_M{}^NO_N=0$. At $\rho=1$ the second term vanishes identically and for the first part at leading order $\mathcal P_M{}^N \bar\nabla_N A^{(S)} \propto \mathcal P_M{}^N  \bar\nabla_N  \rho \propto \mathcal P_M{}^N  n_N $, which also vanishes.  
Therefore
\begin{equation}
\left.
\mathcal P_M{}^N u^P\delta F_{PN}^{(S)}
\right|_{\rho=1}=0.\label{eq:scalar-screen-zero}
\end{equation}
The contribution for the vector gauge field correction to the following spatial projection is
\begin{equation}
\mathcal P_M{}^N\delta F^{(V)}_{uN}
=\mathcal P_M{}^N u^P
\left(
\bar\nabla_P A_N^{(V)}
-\bar\nabla_N A_P^{(V)}
\right).
\label{eq:AV-screen}
\end{equation}
Here also any possible leading derivative of $A_M^{(V)}$ is proportional to $\bar\nabla_M\rho$.
Such terms vanish in eq.~\eqref{eq:AV-screen} because at the horizon
\begin{equation}
u^P\bar\nabla_P\rho=0,
\qquad
\mathcal P_M{}^N\bar\nabla_N\rho=0.
\end{equation}
So at the horizon the following relation holds true
\begin{equation}
\mathcal P_M{}^N\delta F^{(V)}_{uN}|_{(\rho=1)}
=\mathcal O(D^{-1}).
\label{eq:AV-screen-order}
\end{equation}
Combining the leading and subleading contributions
therefore gives
\begin{equation}
\mathcal P_M{}^N F_{uN}
=-\sqrt{2}\,V_M+\mathcal O(D^{-1}).
\label{eq:screen-F-final}
\end{equation}
Using the antisymmetry of $F_{MN}$ the following contraction  gives
\begin{equation}
u^M F_{uM}=u^M u^N F_{NM}=0.\label{U.F}
\end{equation}
Using
$\delta_M{}^N
=\mathcal P_M{}^N+n_Mn^N-u_Mu^N$, along with eq.~\eqref{U.F}
we therefore obtain
\begin{equation}
F_{uM}
=\mathcal P_M{}^N F_{uN}+n_M n^N F_{uN}.
\label{eq:Fu-decomposition}
\end{equation}
Thus eq.~\eqref{eq:screen-F-final} gives
\begin{equation}
F_{uM}
=-\sqrt{2}\,V_M+\mathcal I n_M+\mathcal O(D^{-1}),
\label{eq:Fu-screen-normal}
\end{equation}
where $\mathcal I= n^N F_{uN}$.
Using eqs.~\eqref{eq:ambient-screen-projector},~\eqref{eq:horizon-normalization-use},~and~\eqref{eq:u-normalization}, we get
\begin{equation}
    G^{MN}n_Mn_N=0,
\qquad
G^{MN}n_M\mathcal P_N{}^P F_{uP}=0.\label{Cancellation}
\end{equation}
Also eq.~\eqref{eq:V-ambient} follows $n^MV_M=u^MV_M=0$. Consequently,
\[
O^MV_M=0.
\]
So we can write
\begin{equation}
    G^{MN}V_MV_N
=\eta^{MN}V_MV_N+\mathcal O(D^{-1}).\label{G}
\end{equation}
Now, the induced metric satisfies
\[
\gamma^{ab}\Pi_a{}^M\Pi_b{}^N
=\eta^{MN}-n^Mn^N.
\]
Since $V_a=\Pi_a{}^M V_M$ and $n^MV_M=0$,
it follows that
\begin{equation}
V_aV^a
=\gamma^{ab}V_aV_b
=(\eta^{MN}-n^Mn^N)V_MV_N
=\eta^{MN}V_MV_N
,\label{V}
\end{equation}
where $V^a=\gamma^{ab}V_b$ and $V^M=\eta^{MN}V_N$.
As discussed earlier, the contraction of the Maxwell field tensor in the Einstein-Maxwell equation is with respect to the full spacetime metric $G_{MN}$, using eqs.~\eqref{Cancellation},~\eqref{G}~and~\eqref{V} we find
\begin{align}
F_{uM}F_u{}^M
&=G^{MN}
\bigl(\mathcal P_M{}^P F_{uP}+\mathcal{I}n_M\bigr)
\bigl(\mathcal P_N{}^Q F_{uQ}+\mathcal{I}n_N\bigr)
\nonumber\\
&=2G^{MN}V_MV_N+\mathcal O(D^{-1})
\nonumber\\
&=2\eta^{MN}V_MV_N+\mathcal O(D^{-1})
\nonumber\\
&=2V_aV^a+\mathcal O(D^{-1}),
\label{eq:Fu-square}
\end{align}
Finally, eq.~\eqref{eq:Ruu-Maxwell} which represents the corresponding scalar constraint equation, becomes
\begin{equation}
R_{uu}=V_aV^a+\mathcal O(D^{-1}).
\label{eq:Ruu-V}
\end{equation}
\section{Non-negative divergence of membrane velocity}
\label{sec:secondlaw}
The leading-order Raychaudhuri equation (eq.~\eqref{eq:ray-membrane-variables}) contains an explicit expression of $R_{uu}$. Here we substitute eq.~\eqref{eq:Ruu-V} in~eq.~\eqref{eq:ray-membrane-variables} and get the following relation:
\begin{equation}
\widehat{\nabla}_a u^a
=
\frac{2}{
K(1-Q^2)}
\left(
\sigma_{ab}\sigma^{ab}
+
V_aV^a
\right)
+
\mathcal O(D^{-2}),
\label{eq:div-main}
\end{equation}
where
\begin{equation}
V^2= V_aV^a .
\end{equation}
Both the terms on the right-hand side are non-negative,
\begin{equation}
\sigma^2\geq0,
\qquad
V^2\geq0 .
\end{equation}
Thus, the leading incompressibility condition \cite{Bhattacharyya:2016nhn}
$\widehat{\nabla}_a u^a=0$ acquires a first subleading correction of
order $1/D$.  More importantly, this correction is a sum of non-negative quadratic forms. In the regime considered here, $K(1-Q^2)>0$.
The two non-negative quadratic forms therefore imply
\begin{equation}
\widehat{\nabla}_a u^a\geq0
\label{eq:div-positive}
\end{equation}
at the first non-vanishing order. There are two non-negative contributions to the membrane expansion scalar, one is the shear contribution $\sigma_{ab}\sigma^{ab}$, while the other  $V_aV^a$ arises purely due to the Maxwell field.

\section{Entropy current and the local second law}
\label{sec:area}

For Einstein gravity, the entropy current associated with the large $D$
membrane was constructed in ref.~\cite{Bhattacharyya:2016nhn} from the area form of the event horizon. A similar identification works here for the first
non-vanishing entropy production. For a charged large $D$ membrane, the associated entropy current $J_S^a$ can be written as follows:
\begin{equation}
J_S^a
=\frac\zeta4\,u^a=
\frac14\,u^a
+
\mathcal O(D^{-2}) .
\label{eq:JS-leading}
\end{equation}
The factor $1/4$ is the Bekenstein--Hawking entropy density.
So indeed, the Einstein--Maxwell theory has the usual
Bekenstein--Hawking entropy $S=A/4$, where $A$ is the physical area of the event horizon in the black hole geometry. However, the comparison of the
physical horizon area element with the corresponding membrane spatial area element gives the leading-order matching (as
$
\zeta= \frac{\sqrt{h_{\rm H}}}{\sqrt h}
=
1+\mathcal O(D^{-2})
$).
Hence, eq.~\eqref{eq:JS-leading} represents the actual expression of entropy current at the leading order with a non-negative divergence. As
\begin{equation}
\widehat{\nabla}_a J_S^a
=
\frac14\,\widehat{\nabla}_a u^a
+
\mathcal O(D^{-2}) .
\label{eq:JS-div}
\end{equation}
Using the scalar constraint derived in the previous section,
\begin{equation}
\widehat{\nabla}_a u^a
=
\frac{2}{
K(1-Q^2)}
\left(
\sigma_{ab}\sigma^{ab}
+
V_aV^a
\right)
+
\mathcal O(D^{-2}),
\end{equation}
we obtain
\begin{equation}
\widehat{\nabla}_a J_S^a
=
\frac{
\sigma_{ab}\sigma^{ab}
+
V_aV^a
}{
2K(1-Q^2)
}
+
\mathcal O(D^{-2}) .
\label{eq:entropy-production}
\end{equation}
With
\begin{equation}
    V_a
=P_a{}^b\widehat\nabla_bQ+Qu\cdot\widehat\nabla u_a.
\end{equation}
Since $K(1-Q^2)>0$ in the regime considered here, and all the quantities $\sigma_{ab}\sigma^{ab}$, $K$ and $V_aV^a$
are non-negative, so at the leading non-vanishing order we have
\begin{equation}
\widehat{\nabla}_a J_S^a\geq0 .
\label{eq:local-second-law}
\end{equation}
Thus, the non-negative divergence of the membrane velocity has the direct interpretation of local entropy production. In addition to
the familiar shear contribution of the neutral membrane
\cite{Dandekar:2016fvw,Bhattacharyya:2016nhn,Saha:2020zho}, the charged membrane
contains the additional contribution $V_aV^a$ to the entropy production generated by the Maxwell field.
\section{Limiting conditions}
\label{sec:checks}

\subsection{Neutral limit}

In the neutral limit, we have
\begin{equation}
 Q=0,
 \qquad
 \widehat{\nabla}_{a}Q=0.
\end{equation}
Then using eq.~\eqref{eq:V-def} we have
\begin{equation}
 V_{a}=0
\end{equation}
and eq.~\eqref{eq:div-main} reduces to
\begin{equation}
 \widehat{\nabla}_{a}u^{a}
 =
 \frac{2\sigma_{ab}\sigma^{ab}}{K}
 +
 \mathcal{O}(D^{-2}).
 \label{eq:neutral-div}
\end{equation}
To compare directly with the convention used in
refs.~\cite{Dandekar:2016fvw,Bhattacharyya:2016nhn},
we define the following quantity
\begin{equation}
 S_{ab}
 =
 P_{a}{}^{c}P_{b}{}^{d}
 \left(
 \widehat{\nabla}_{c}u_{d}
 +
 \widehat{\nabla}_{d}u_{c}
 \right).
 \label{eq:S-unhalved}
\end{equation}
Since
$\widehat{\nabla}\!\cdot u=\mathcal{O}(D^{-1})$,
\begin{equation}
 S_{ab}
 =
 2\sigma_{ab}
 +
 \mathcal{O}(D^{-1}).
\end{equation}
Consequently,
\begin{equation}
 \widehat{\nabla}\!\cdot u
 =
 \frac{
 S_{ab}S^{ab}
 }{
 2K
 }
 +
 \mathcal{O}(D^{-2}),
 \end{equation}
which is the known neutral first-subleading result
\cite{Dandekar:2016fvw}.
Using eq.~\eqref{eq:JS-leading} we have
\begin{equation}
 \widehat{\nabla}_{a}J_{S}^{a}
 =
 \frac{
 S_{ab}S^{ab}
 }{
 8K
 }
 +
 \mathcal{O}(D^{-2}).
\end{equation}
The above expression is in agreement with the entropy-production formula of ref.~\cite{Bhattacharyya:2016nhn}. 
\subsection{Regime of applicability}
The charge field $Q(x)$ may vary along the membrane,
subject to the condition~\eqref{eq:charge-domain}.
The quantity $1-Q(x)^2$ therefore remains positive
and of order unity in the large $D$ limit. This
scaling keeps the leading non-affinity
$\kappa=K(1-Q^2)/2+\mathcal O(1)$ of order $D$,
so that the shear and Maxwell contributions determine
the horizon expansion, and hence the membrane entropy
production, at order $1/D$. The result thus applies
locally to dynamical charged membranes within this
regime.
\section{Summary and outlook}
\label{sec:discussion}
In summary, this work constructs the entropy current of the charged large $D$ membrane paradigm at the leading order. The timelike membrane velocity corresponds to a null generator of the black-hole event horizon, which obeys the null Raychaudhuri equation. This equation constrains the null--null component of the Ricci tensor. Using the Einstein--Maxwell equations in the large $D$ limit, we evaluate this component and substitute it into the null Raychaudhuri equation to obtain the first non-vanishing contribution to horizon expansion in the black-hole geometry. Translating this result to the membrane picture gives the membrane expansion scalar, whose leading contribution is non-negative, as in the uncharged case. Finally, using this result, we construct the membrane entropy current (eq.~\eqref{eq:JS-leading}) and its non-negative divergence (eq.~\eqref{eq:entropy-production}), which captures the local entropy production of the charged large $D$ membrane paradigm. 
The resulting entropy-production equation therefore
provides an explicit realization of the local second law in the charged large $D$ membrane description, with the electromagnetic field supplying an additional non-negative contribution beyond the gravitational shear term.

Looking forward, an interesting question is how the entropy-production formula changes when the local charge profile approaches
$Q(x)^2=1$. Addressing this question would involve
taking the charge limit $(\epsilon\to0)$ together with the large $D$
limit $(D\to\infty)$ and determining the resulting balance in the
horizon scalar constraint. It would also be interesting
to extend the calculation to charged Gauss--Bonnet
gravity and examine the contribution of the
higher-curvature terms to the membrane entropy current.

\appendix
\section{Large D scaling of $\ln\zeta$ }\label{Some}

For the second term in
eq.~\eqref{eq:zeta-determinant-expansion},
using eq.~\eqref{eq:screen-metric-correction},
we obtain
\begin{align}
&h^{ik}\delta h_{kj}h^{jl}\delta h_{li}
\nonumber\\
&=h^{ik}
\left(H^{(T)}_{kj}+H^{(\mathrm{Tr})}h_{kj}
+\mathcal O(D^{-2})\right)
h^{jl}
\left(H^{(T)}_{li}+H^{(\mathrm{Tr})}h_{li}
+\mathcal O(D^{-2})\right)
\nonumber\\
&=h^{ik}H^{(T)}_{kj}h^{jl}H^{(T)}_{li}
+2H^{(\mathrm{Tr})}h^{ij}H^{(T)}_{ij}
+(D-2)(H^{(\mathrm{Tr})})^2
+\mathcal O(D^{-3})
\nonumber\\
&=H^{(T)}_{ij}H_{(T)}^{ij}
+\mathcal O(D^{-3}),
\end{align}
where we used
$h^{ik}H^{(T)}_{kj}=H^{(T)i}{}_{j}$. Under the $SO(D-p-2)$ symmetry, the spatial indices
split into $p+1$ non-spherical directions
$\alpha,\beta$ and $D-p-3$ spherical directions
$I,J$. Rotational invariance implies
$H^{(T)}_{\alpha I}=0$ and
using the tracelessness condition
\begin{equation}
h^{ij}H^{(T)}_{ij}
=h^{\alpha\beta}H^{(T)}_{\alpha\beta}
+h^{IJ}H^{(T)}_{IJ}=0,
\end{equation}
we obtain
\begin{equation}
H^{(T)}_{IJ}
=-\frac{h_{IJ}}{D-p-3}
h^{\alpha\beta}H^{(T)}_{\alpha\beta}.
\end{equation}
Consequently,
\begin{align}
H^{(T)}_{ij}H_{(T)}^{ij}
&=H^{(T)}_{\alpha\beta}H_{(T)}^{\alpha\beta}
+H^{(T)}_{IJ}H_{(T)}^{IJ}
\nonumber\\
&=H^{(T)}_{\alpha\beta}H_{(T)}^{\alpha\beta}
+\frac{(h^{\alpha\beta}H^{(T)}_{\alpha\beta})^2}
{D-p-3}.
\end{align}
Since $p$ is fixed, the number of non-spherical
directions is $\mathcal O(1)$. Using
$H^{(T)}_{\alpha\beta}=\mathcal O(D^{-1})$,
we have
\begin{align}
H^{(T)}_{\alpha\beta}H_{(T)}^{\alpha\beta}
&=\mathcal O(D^{-2}),\\
h^{\alpha\beta}H^{(T)}_{\alpha\beta}
&=\mathcal O(D^{-1}),\\
\frac{(h^{\alpha\beta}H^{(T)}_{\alpha\beta})^2}
{D-p-3}
&=\mathcal O(D^{-3}).
\end{align}
Therefore,
\begin{equation}
H^{(T)}_{ij}H_{(T)}^{ij}
=\mathcal O(D^{-2}).
\end{equation}
Consequently,
\begin{equation}
h^{ik}\delta h_{kj}h^{jl}\delta h_{li}
=\mathcal O(D^{-2}).
\end{equation}

Similarly the other components (cubic and higher-order terms of $\delta h_{kj}$) in eq.~\eqref{eq:zeta-determinant-expansion}
contributes to order higher than $\mathcal O(D^{-2})$, leading to
\begin{equation}
    \ln\zeta=\mathcal{O}(D^{-2}).
\end{equation}

\section*{Acknowledgments}
The authors express their gratitude to the citizens of India for their continued support of scientific research. S.H. is supported by an Institute Fellowship from the Indian Institute of Technology (ISM) Dhanbad. We also thank the Department of Physics at IIT (ISM) Dhanbad for providing essential research facilities.

\printbibliography
\end{document}